\documentclass{jors}

\usepackage{listings}
\usepackage{graphicx}
\definecolor{mygray}{gray}{0.6}

\usepackage{color}
\usepackage{float}

\definecolor{mygreen}{rgb}{0,0.6,0}
\definecolor{mygray}{rgb}{0.5,0.5,0.5}
\definecolor{mymauve}{rgb}{0.58,0,0.82}

\usepackage{hyperref}

\begin{document}

{\bf Software paper for submission to the Journal of Open Research Software} \\



\rule{\textwidth}{1pt}

\section*{(1) Overview}
\label{sec:overview}

\vspace{0.5cm}

\section*{Title}


Turbulucid: A Python Package for Post-Processing of Fluid Flow Simulations

\section*{Paper Authors}


Mukha, Timofey

\section*{Paper Author Roles and Affiliations}

Department of Information Technology, Uppsala University, Sweden.

\section*{Abstract}


A Python package for post-processing of plane two-dimensional data from computational fluid dynamics simulations is presented.
The package, called \texttt{turbulucid}, provides means for scripted, reproducible analysis of large simulation campaigns and includes routines for both data extraction and visualization.
For the former, the  Visualization Toolkit (VTK) is used, allowing for post-processing of simulations performed on unstructured meshes.
For visualization, several \texttt{matplotlib}-based functions for creating highly customizable, publication-quality plots are provided.
To demonstrate \texttt{turbulucid}'s functionality it is here applied to post-processing a simulation of a flow over a backward-facing step.
The implementation and architecture of the package are also discussed, as well as its reuse potential.

\section*{Keywords}

visualization; computational fluid dynamics; data analysis; post-processing;

\section*{Introduction}


Performing a computational fluid dynamics (CFD) simulation can generally be divided into three stages: pre-processing, setting up and running the solver, and post-processing.
The pre-processing stage consists of defining the computational domain and discretizing it with a mesh. 
Next, the solver is configured to reflect the physics of the problem and run in order to obtain a solution, which is, in general, a three-dimensional time-dependent dataset.
At the post-processing stage, the produced solution is analysed.
This includes both extracting relevant data (e.g. pressure or velocity values at certain locations) and visualizing it with different types of plots.

Focusing on the post-processing stage, there is a large variety of tools providing associated functionality.
Indeed, most CFD software packages, while concentrating on solvers, usually also provide means for conducting basic post-processing.
For example, routines for extracting data along a cut-plane or a line are commonly present, as well as the possibility for producing different types of plots (scatter plots, vector plots, streamline plots etc).
Specialized software for post-processing exists as well, and typically provides a richer functionality, better performance, and higher-quality rendering.
However, the existing solutions tend to:
\begin{itemize}
    \item Focus on working with large unstructured three-dimensional data.
    \item Provide limited options for customizing the plots (e.g. fonts, sizes and styles of different plot elements, etc).
    \item Excel in interactivity, but not reproducibility (e.g. quickly producing the same plot for 10 different datasets).
\end{itemize}
Indeed, the properties above reflect the average post-processing needs of the users.
Nevertheless, there are many situations when a different sort of tool is required.
While the absolute majority of CFD datasets are in fact three-dimensional, and many are also time-dependent, it is hard to analyse such data as is. 
Commonly, time-averaging is applied along with plane- or line-data extraction, and the actual analysis and visualization are thus performed on data of lower dimension.
Fine-grain plot customization may not be needed in most applications but is an absolute must when producing figures for publications.
Finally, interactivity is important when looking at a single case for the first time, but being able to quickly reproduce the analysis or apply it to a new dataset becomes increasingly important in larger simulation campaigns.

It would, therefore, be beneficial to have a complementary post-processing tool that excels at performing easily reproducible analyses of two-dimensional datasets as well as producing high-quality customizable visualizations.
The focus of this work is presenting such a tool, namely, a Python package called \texttt{turbulucid}.
By combining data manipulation functionality provided by the Visualization Toolkit (VTK)~\cite{vtkbook} and plotting routines available in \texttt{matplotlib}~\cite{matplotlib}, \texttt{turbulucid} allows to produce publication-quality plots of several types and provides functionality for analysing the dataset with a set of data extraction routines.
Using the package, the post-processing can be scripted in Python using the provided objects and functions.
This makes it easy to analyse a large set of simulations in a structured way or quickly apply an existing analysis to a new case.
Additionally, users have the possibility to use any of the numerous packages available in the Python ecosystem in their analysis.
   
While the package works exclusively with planar two-dimensional datasets, no assumption is made regarding the topology of the computational mesh,  meaning that any mesh consisting of polygonal cells can be used.
The data itself is assumed to be stored in a format for unstructured meshes defined by VTK, but the package is designed to be easily extendible to read data in other formats.
Scalar, vector and tensorial quantities are supported.

The discussion of the features of \texttt{turbulucid}, as well as its design and implementation, is continued in the sections below.
Examples of applying \texttt{turbulucid} to post-processing of simulations of various flows can be found in~\cite{Rezaeiravesh2016, Liefvendahl2016, Liefvendahl2017a}.
This includes flat-plate turbulent boundary layer flow, flow around a ship hull, and also flow around a submarine-like axisymmetric body.
Here, the functionality of \texttt{turbulucid} is demonstrated by applying it to post-processing of a simulation of a flow over a backward-facing step.

\section*{Implementation and architecture}

As it was mentioned in the introduction, \texttt{turbulucid} uses Python bindings for VTK to handle the unstructured mesh and \texttt{matplotlib}'s plotting routines for producing plots.
It is important to note that the user is never exposed to VTK objects, therefore familiarity with VTK's API is not a prerequisite for using \texttt{turbulucid}.
All data is instead returned to the user as \texttt{numpy}~\cite{numpy} arrays.

By contrast, the constructed plots are returned to the user as objects of the appropriate type defined by \texttt{matplotlib} (e.g a \texttt{StreamPlotSet} object for a streamline plot.)
This allows for customizing the created plots.

To open a dataset, the user has to create a \texttt{Case} object.
The path to the dataset is passed to the constructor.
Once created, the methods and attributes of the \texttt{Case} object provide access to the dataset.
For example, the \texttt{\_\_getitem\_\_} operator is overloaded to return the values of a field present in the dataset, given the field's name.

To make \texttt{turbulucid} easily extendible to work with datasets saved in various formats, a separate hierarchy of classes responsible for reading the data is present.
Currently, readers for legacy and XML VTK data files (extensions .vtk and .vtu, respectively) are implemented, the corresponding classes being \texttt{LegacyReader} and \texttt{XMLReader}.
Both implemented readers are derived from \texttt{Reader}, which serves as a base abstract class.
The \texttt{Case} class determines which \texttt{Reader}-class should be used based on the extension of the file the dataset is stored in.

Internally, the dataset is stored as \texttt{vtkPolyData}. 
Note that this and other VTK formats support data of two types: cell and point.
The former associates a value with each polygonal cell whereas the latter associates a value with each mesh-node.
In \texttt{turbulucid}, the fields are assumed to be stored as cell data.
If point data is stored instead, the readers perform linear interpolation in order to produce corresponding cell data.

To reduce the amount of boilerplate code that has to be written by the user, object-oriented programming is not used for implementing the plotting and data extraction features.
Instead, they are implemented as functions, which commonly require a \texttt{Case} object as an input parameter.
The provided functions and their purpose are summarized in Tables~\ref{tab:plot} and \ref{tab:data}.

\begin{table}[htp!]
\centering
\caption{Plot functions available in \texttt{turbulucid}.}
\label{my-label}
\begin{tabular}{lp{10cm}}
\textbf{Name}                    & \textbf{Produced plot} \\ \hline
\texttt{plot\_field}             & Each cell is colored with the corresponding value of a given scalar field.           \\
\texttt{plot\_vectors}           & Arrows showing the magnitude and direction of a  vector field.   \\
\texttt{plot\_streamlines}       & Streamlines following a vector field.                          \\
\texttt{plot\_boundaries}        & Lines showing the boundaries of the geometry.                    \\
\texttt{add\_colorbar}           & Adds a colorbar.                                                                             
\end{tabular}
\label{tab:plot}
\end{table}

\begin{table}[htp!]
\centering
\caption{Data extraction functions available in \texttt{turbulucid}.}
\label{my-label}
\begin{tabular}{lp{10cm}}
\textbf{Name}                    & \textbf{Purpose} \\ \hline
\texttt{profile\_along\_line}    & Extract data along a line.\\
\texttt{sample\_by\_plane}       & Re-sample the dataset using a Cartesian grid.\\
\texttt{dist}                    & Compute distances from centres of boundary-adjacent  cells to the boundary.                          \\
\texttt{normals}                 & Compute unit outward normals to every edge of a given boundary.                    \\
\texttt{tangents}                & Compute unit tangent vectors to every edge of a given boundary.                                                                             
\end{tabular}
\label{tab:data}
\end{table}

The package is documented using \texttt{numpy}-style docstrings.
The \texttt{Sphinx} package is used to compile them into \texttt{html}\footnote{See \texttt{https://timofeymukha.github.io/turbulucid}}.

\section*{Demonstration of functionality}
In this section \texttt{turbulucid} is used to post-process results from a simulation of a flow over a backward-facing step (BFS)\footnote{The simulation results were provided by Saleh Rezaeiravesh from Uppsala University through personal communication.}.
The goal is to demonstrate the quality of some of the plot types \texttt{turbulucid} can be used to produce.
The analysis of the flow as such is therefore kept at a superficial level.
For completeness, it is noted that the simulations results were obtained by conducting a large-eddy simulation of the flow using the open-source CFD software \texttt{OpenFOAM}~\cite{Weller1998}.
The unknowns were averaged in time in the course of the simulation and then also across the statistically homogeneous spanwise direction, thus producing a two-dimensional dataset.

\begin{figure}[H]
    \includegraphics[scale=0.9, trim=0.5cm 3.6cm 0.4cm 4cm, clip=true]{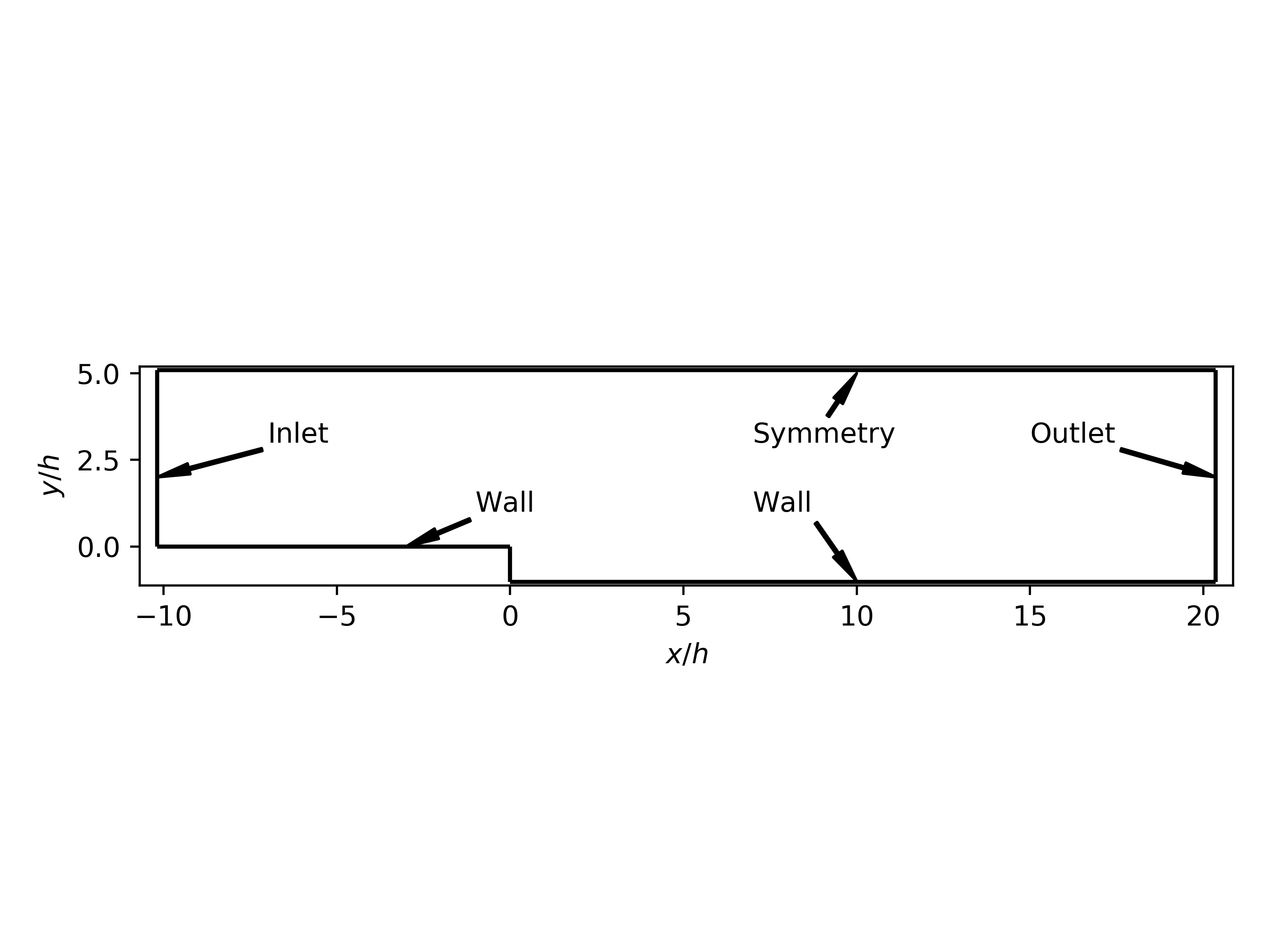}
    \caption{Computational domain of the BFS simulation.}
    \label{fig:boundaries}
\end{figure}

To show the computational domain, the function \texttt{plot\_boundaries} can be used.
It is possible to scale the $x$ and $y$ axis.
In the case of the BFS, it is common to use the step-height, $h$, as a scaling parameter.
We can also use \texttt{matplotlib} to add annotations to the figure to indicate what boundary conditions are used, as well as add axes labels, see Figure~\ref{fig:boundaries}.
This example clearly illustrates how \texttt{turbulucid} seamlessly integrates with \texttt{matplotlib} allowing the user to take full advantage of this library.

\begin{figure}[H]
    \centering
    \includegraphics[scale=0.9, trim=0.5cm 3.6cm 0cm 4cm, clip=true]{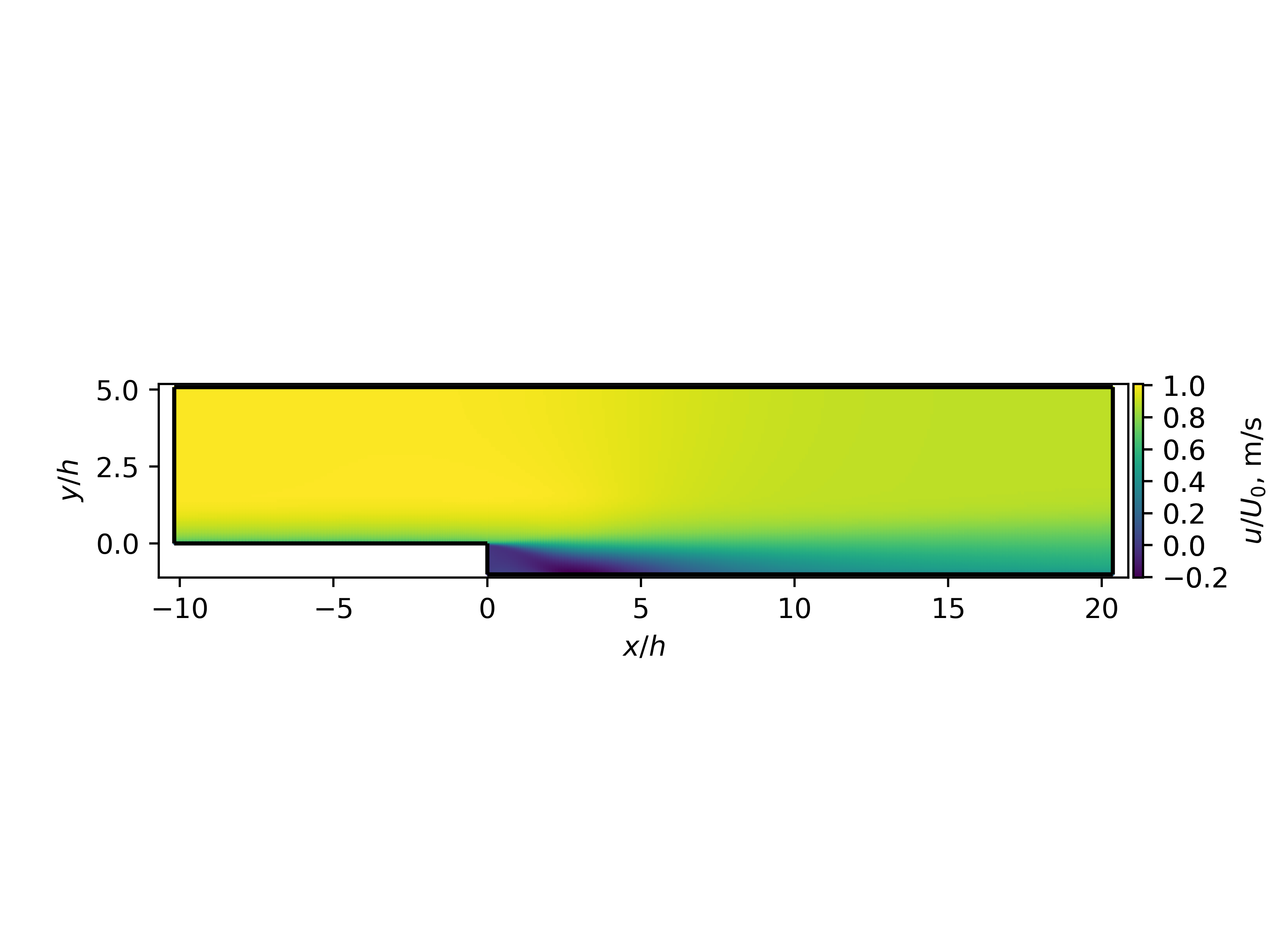}
    \caption{Distribution of the wall-parallel velocity across the domain. Values normalized with a reference velocity, $U_0$.}
    \label{fig:u}
\end{figure}

To get a good qualitative understanding of the flow, let us plot the distribution of the wall-parallel component of velocity across the geometry.
To this end, the function \texttt{plot\_field} can be used, and \texttt{add\_colorbar} can be used to add a colorbar to the plot, see Figure~\ref{fig:u}.

The code used to produce Figure~\ref{fig:u} is given in Listing~\ref{lst:fig2}.
First, the data is opened by creating a \texttt{Case} object.
Then the step-height, \texttt{h}, is defined, followed by a call to the \texttt{plot\_field} function that creates the plot.
Several arguments are passed to \texttt{plot\_field}.
The created \texttt{Case} object is the first argument.
The second argument is a \texttt{numpy} array of values to be plotted.
The array is retrieved from the \texttt{Case} object by passing the name of the desired field (here \texttt{UMean}, i.e the mean velocity) to the \texttt{\_\_getitem\_\_} operator.
Finally, the keyword arguments \texttt{scaleX} and \texttt{scaleY} are set to \texttt{h} to scale the axes of the plot with the step-height.
The created plot-object is assigned to a variable, \texttt{f}.
A colorbar object is then created using \texttt{add\_colorbar} and \texttt{f}. The object is then manipulated to set the correct colorbar label.
Finally, $x$- and $y$-labels are defined using standard functions from \texttt{matplotlib.pyplot}, here imported as \texttt{plt}.
\newpage

\begin{lstlisting}[caption=Code snippet used to produce Figure 2., label=lst:fig2]
case = Case("path/to/data")
h = 0.0094318
f = plot_field(case, case["UMean"][:,0], scaleX=h, scaleY=h)
cbar = add_colorbar(f)
cbar.ax.set_ylabel(r"$u/U_0$, m/s")
plt.xlabel(r"$x/h$")
plt.ylabel(r"$y/h$")
\end{lstlisting}

\begin{figure}[H]
    \centering
    \includegraphics[scale=0.9, trim=0.4cm 3.2cm 0.4cm 4cm, clip=true]{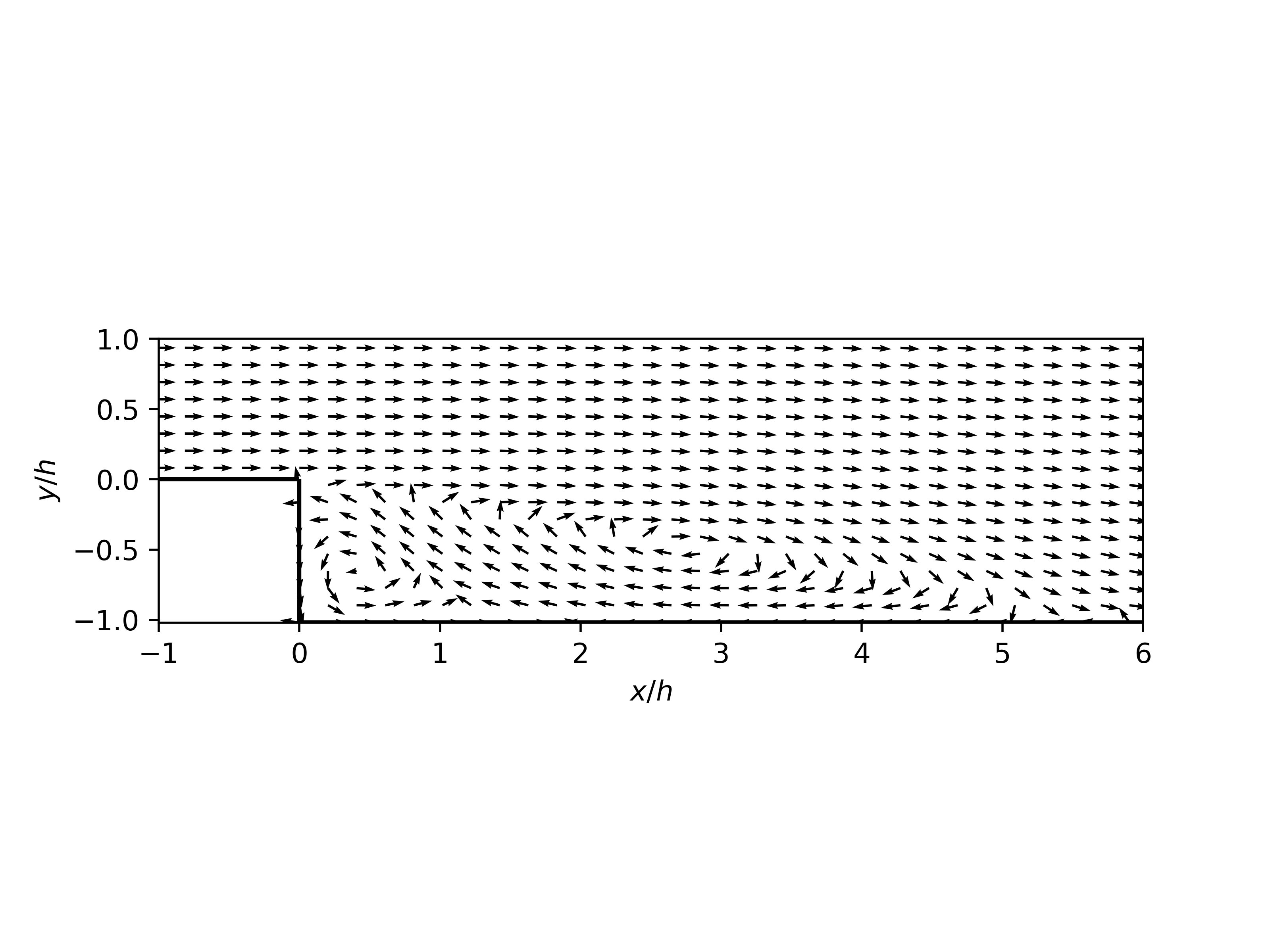}
    \caption{Velocity vectors in and around the recirculation region.}
    \label{fig:vec}
\end{figure}

It can be seen in Figure~\ref{fig:u} that a boundary layer approaches the step from the left, separates, and reattaches at $x/h \approx 6$.
A recirculation region is formed directly downstream of the step.
To investigate this region further, a vector plot can be created using the \texttt{plot\_vectors} function, see Figure~\ref{fig:vec}.
Two recirculation bubbles can be observed: the main, larger, bubble and a secondary one in the corner directly downstream of the step.

\begin{figure}[htp!]
    \centering
    \includegraphics[scale=0.9, trim=0.4cm 3.2cm 0.4cm 4cm, clip=true]{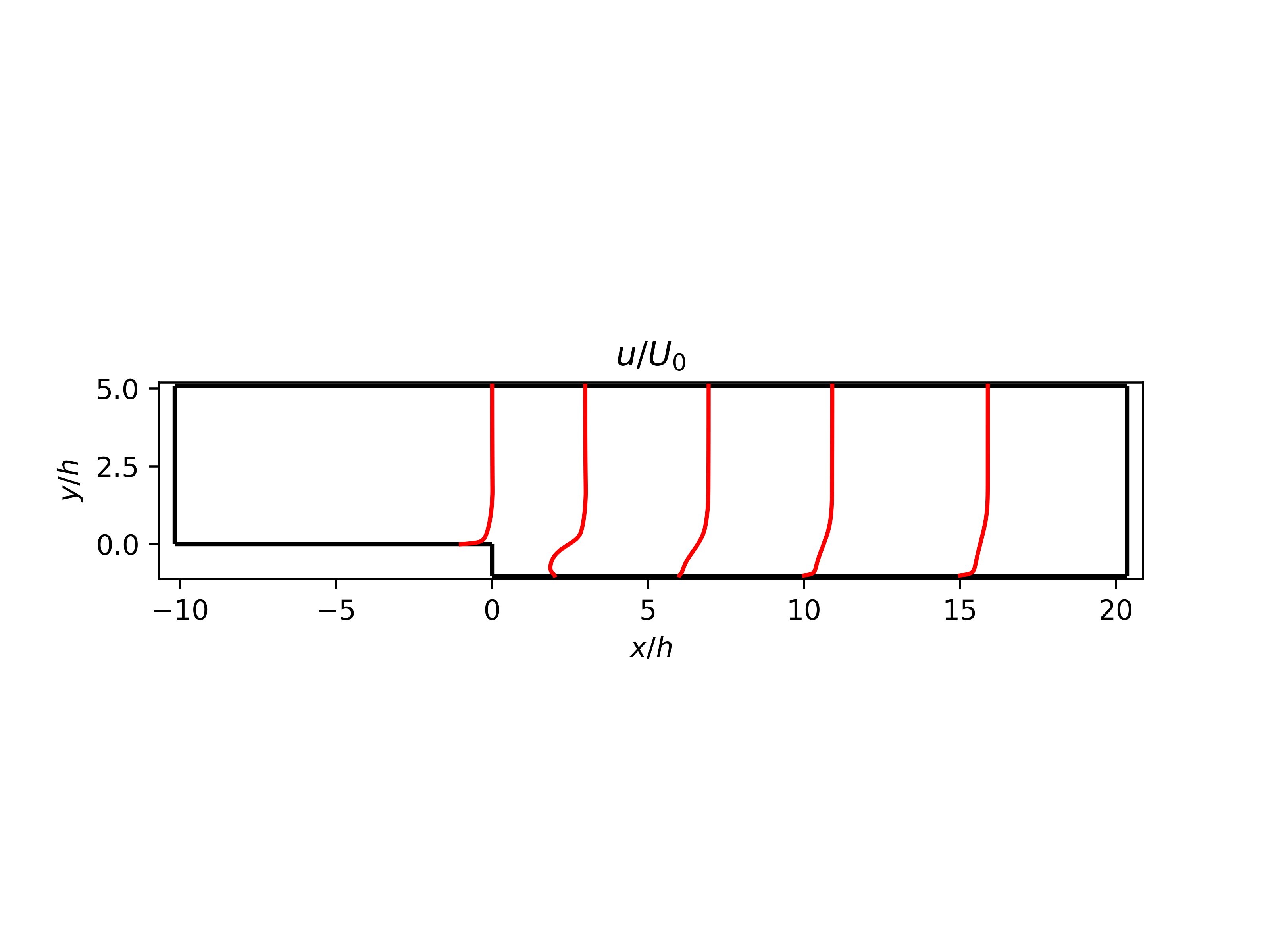}
    \caption{Profiles of wall-parallel velocity at different streamwise locations. Values normalized with a reference velocity, $U_0$.}
    \label{fig:profiles}
\end{figure}

Profile plots are commonly used to compare the solution  with reference data, obtained computationally or experimentally.
For the BFS in particular, profiles of the $x$ component of velocity as a function of $y$, at different streamwise locations, can be considered.
To extract data along a line the \texttt{profile\_along\_line} function can be used.
It is then possible to combine \texttt{plot\_boundaries} with \texttt{matplotlib}'s \texttt{plot} function to embed line-plots into the geometry of the computational domain, see Figure~\ref{fig:profiles}.

The inflection of the velocity profile is clearly seen in the separation region, at $x/h=2$, whereas at $x/h \approx 6$ the inflection vanishes, indicating reattachment.
A recovery of a canonical turbulent boundary layer profile is observed downstream.

\section*{Quality control}

The framework \texttt{pytest} is used to test the implemented functionality with unit tests.
Travis CI is used to automatically test installing the package and running all the tests, using both Python 3 and Python 2. 

\section*{(2) Availability}
\vspace{0.5cm}
\section*{Operating system}
The package is expected to run on any operating systems, which are supported by all the dependencies (see below).
This includes but is not limited to modern Linux distributions and Windows.

\section*{Programming language}

\texttt{Turbulucid} is written in Python 3, but is compatible and tested with Python 2 as well.

\section*{Additional system requirements}
None.

\section*{Dependencies}

The following Python packages: \texttt{numpy}, \texttt{matplotlib}, \texttt{scipy}.
VTK version 7.0.0 or higher, and the associated Python bindings.
The \texttt{Sphinx} package is needed to build the documentation.
\section*{List of contributors}

\begin{itemize}
\item Timofey Mukha, Uppsala University. Development, testing, writing documentation.
\item Saleh Rezaeiravesh, Uppsala University. Validation of functionality.

\item Mattias Liefvendahl, Uppsala University and Swedish Defence Research Agency (FOI). Validation of functionality.
\end{itemize}

\section*{Software location:}



{\bf Code repository} Github

\begin{description}[noitemsep,topsep=0pt]
	\item[Name:] \texttt{turbulucid}
	\item[Persistent identifier:] \texttt{https://github.com/timofeymukha/turbulucid}
	\item[Licence:] GNU GPL version 3
	\item[Date published:] 02/03/2016
\end{description}

\section*{Language}

English

\section*{(3) Reuse potential}


\texttt{Turbulucid} can be useful to all engineers and researchers working with computational fluid dynamics.
In particular, when there is need for producing a publication-quality plot or performing an easily reproducible scripted analysis of a simulation campaign.

The package can be used directly with any CFD-solver that supports extracting cut-plane data in VTK format.
Otherwise, the data should first be converted into the appropriate format, e.g. using the VTK API.
The \texttt{turbulucid} package itself can also be extended to read in data stored in a different format and apply appropriate conversion routines on the fly.
Such contributions are most welcome, and anyone willing to extend \texttt{turbulucid} in this or any other way is encouraged to contact the author or open an issue in the Github repository.

A \textit{readme} file, including installation instructions, is provided with the software.
Additionally, a tutorial in form of a Jupyter notebook~\cite{jupyter} is provided, demonstrating most of the functionality of the package. 
While further support cannot be guaranteed, the author will do his best to provide aid to users.
Github issues can be used for asking for help.

\section*{Acknowledgements}

The incentive to create \texttt{turbulucid} came from the need to post-process simulations conducted using computing resources provided by the Swedish National Infrastructure for Computing (SNIC).
Therefore SNIC and, in particular, the PDC Centre for High Performance Computing (PDC-HPC) are gratefully acknowledged.

\section*{Funding statement}
 The work was supported by Grant No 621-2012-3721 from the Swedish Research Council.

\section*{Competing interests}

The author has no competing interests to declare.



\bibliographystyle{plain}
\bibliography{library}



%
%
%

\end{document}